# The Radio Emission from the Ultra-luminous Far-infrared Galaxy NGC 6240

Edward J. M. Colbert and Andrew S. Wilson[1]

Department of Astronomy, University of Maryland, College Park, MD 20742

and

Jonathan Bland-Hawthorn

Anglo-Australian Observatory, P. O. Box 296, Epping, NSW 2121, Australia

[1]also Space Telescope Science Institute, 3700 San Martin Drive, Baltimore, MD 21218

## ABSTRACT

We present new radio observations of the "prototypical" ultra-luminous far-infrared galaxy NGC 6240, obtained using the VLA at $\lambda = 20$ cm in 'B' configuration and at $\lambda = 3.6$ cm in 'A' configuration. These data, along with those from four previous VLA observations, are used to perform a comprehensive study of the radio emission from NGC 6240. Approximately 70% ($\sim$3 $\times$ $10^{23}$ W $Hz^{-1}$) of the total radio power at 20 cm originates from the nuclear region ($\lesssim$1.5 kpc), of which half is emitted by two unresolved ($R \lesssim 36$ pc) cores and half by a diffuse component. The radio spectrum of the nuclear emission is relatively flat ($\alpha \approx 0.6$; $S_\nu \propto \nu^{-\alpha}$). The supernova rate required to power the diffuse component is consistent with that predicted by the stellar evolution models of Rieke et al. (1985). If the radio emission from the two compact cores is powered by supernova remnants, then either the remnants overlap and form hot bubbles in the cores, or they are very young ($\lesssim$100 yr). Nearly all of the other 30% of the total radio power comes from an "arm-like" region extending westward from the nuclear region. The western arm emission has a steep spectrum ($\alpha \approx 1.0$), suggestive of aging effects from synchrotron or inverse-Compton losses, and is not correlated with starlight; we suggest that it is synchrotron emission from a shell of material driven by a galactic superwind. Inverse-Compton scattering of far-infrared photons in the radio sources is expected to produce an X-ray flux of $\sim$2–6 $\times$ $10^{-14}$ erg $s^{-1}$ $cm^{-2}$ in the 2–10 keV band. No significant radio emission is detected from or near the possible ultra-massive "dark core" hypothesized by Bland-Hawthorn, Wilson & Tully (1991).

## 1. Introduction

NGC 6240 was identified as an exceptional system as early as 1961, when Zwicky, Herzog & Wild remarked that a "multiple collision" may explain the many distorted tails (Figure 1) extending into the outskirts of the galaxy. Indeed, numerical simulations by Toomre & Toomre (1972) demonstrated that a close encounter between two galaxy disks of comparable mass can produce such strong and extended tails. Fosbury & Wall (1979) again remarked on the "spectacularly disturbed morphology" of NGC 6240 and interpreted the optical and radio emission as being due to collisions of the gas from two galaxies. In 1983, Fried & Schulz acquired optical images of the nuclear region, which revealed two bright peaks separated by 1.8″ (∼900[1] pc), possibly "remnant nuclei" from two progenitor galaxies. Clearly the interpretation of NGC 6240 as a merging system enjoys wide acceptance.

When a strong IRAS source was identified within a few arcseconds of the galaxy nucleus (Wright, Joseph & Meikle 1984), NGC 6240 joined the newly discovered class of galaxies for which ∼99% of the bolometric luminosity emerges in the far-infrared (FIR; Soifer et al. 1984a). The FIR luminosity (1−1000$\mu$m) of $\sim 4.7 \times 10^{11}$ $L_\odot$ (using the method of Helou et al. 1988) from NGC 6240 distinguishes it as an "ultra-luminous" FIR galaxy. A significant percentage (12−25%) of luminous FIR galaxies are known to be interacting systems (Soifer et al. 1984b), which suggests that merger events and FIR excesses are somehow connected. NGC 6240 became an extremely popular galaxy to study as explanations for the origin of the FIR emission were sought, and has become known as a "prototypical" ultra-luminous FIR galaxy.

The most common explanation for the FIR excess is a luminous starburst in a dusty medium (Rieke et al. 1985; Heckman, Armus & Miley 1987; Lester, Harvey & Carr 1988; Smith, Aitken & Roche 1989; Armus, Heckman & Miley 1990). The FIR emission is interpreted as thermal emission from dust heated by hot stars from the starburst, which, in the case of NGC 6240, may have been triggered by an interaction.

[1]We assume a distance of 100 Mpc and a Hubble constant of $H_o$ = 75 km s$^{-1}$ Mpc$^{-1}$ throughout this paper, which corresponds to a scale of ≈0.5 kpc/″.

A second popular explanation is an active galactic nucleus (AGN) surrounded by dust. In this scenario, the dust is heated by ultraviolet (UV) radiation from an active nucleus rather than by hot stars (De Poy, Becklin & Wynn-Williams 1986; Andreasyan & Khachikyan 1987; Lester et al. 1988; Keel 1990). The bolometric luminosity and space density of ultra-luminous FIR galaxies are similar to those of quasars. For this reason, it has been suggested that there is an evolutionary link between the two classes (cf. Rieke 1992). However, further observational work is needed to determine whether ultra-luminous FIR galaxies contain "hidden quasars."

Other interpretations include dust heated by UV radiation from collisions between molecular clouds (Harwit et al. 1987) or dust heated by an older stellar population (Thronson et al. 1990).

No compelling evidence has yet been presented for a definitive interpretation of the FIR excess in luminous FIR galaxies. Studies of NGC 6240 at optical and near-infrared (NIR) wavelengths have yielded ambiguous results, in part because of uncertainty in the distribution of obscuring dust. At infrared wavelengths of a few $\mu$m, the reddening is less severe, although not negligible (cf. Schulz et al. 1993). The flux in the hydrogen recombination lines in the NIR reported by De Poy et al. (1986) is considerably lower than that which would be expected from a luminous starburst. However, different estimates of the reddening yield different values for the rate of emission of ionizing photons, and thus allow different interpretations (e.g. Rieke et al. 1985 and De Poy et al. 1986).

The IRAS flux from spiral and starburst galaxies is known to be tightly correlated with the non-thermal 1.4 GHz radio flux (e.g. Helou, Soifer & Rowan-Robinson 1985), indicating a close connection between the emission processes in the two wavebands. This radio-FIR correlation may, in principle, be accounted for by the starburst scenario: the FIR emission comes from dust heated by hot stars, and the non-thermal radio emission results from starburst-generated supernovae (SNe), supernova remnants (SNRs) and relativistic electrons which have escaped from the SNRs. In contrast to the galaxies that follow this radio-FIR correlation, Seyfert galaxies tend to have stronger radio flux for a given FIR flux by up to one order of magnitude (e.g. Wilson 1988). Helou et al. (1985) have found that, for spirals and starbursts, the $q$-parameter, defined as

$$q = log\left\{\frac{F_{FIR}/3.75 \times 10^{12}\ \mathrm{Hz}}{S_\nu(1.4\ \mathrm{GHz})}\right\},$$

where $F_{FIR}$ is the IRAS FIR parameter (see e.g. Fullmer & Lonsdale 1989) and $S_\nu$ is the radio flux, has a mean value of 2.14 and a dispersion of 0.14. The value of $q$ for NGC 6240 is 1.88. Although this relatively low value of $q$ indicates that the radio flux may contain a component in addition to that from the starburst, the difference from the mean value is not sufficient evidence for the existence of an active nucleus.

The absence of obscuring effects at radio wavelengths make radio observations of luminous FIR galaxies an attractive probe of the putative starburst or AGN. Previous workers have used individual radio maps to discuss possible starburst and AGN components in NGC 6240. Most of the radio emission in the 20 cm map (resolution $\sim2''$) by Condon et al. (1982) is concentrated in the galaxy nucleus. Their higher-resolution ($\sim0.6''$) 6 cm map resolves the nuclear emission into a double-peaked structure similar to that observed at optical and NIR wavelengths. Condon et al. associated the radio emission with a burst of star formation, possibly triggered by an interaction. Carral, Turner & Ho (1990) obtained a higher-resolution ($\sim0.15''$) image of the nuclear region at 2 cm and found 60% of the total flux in their image to be in two unresolved ($R \lesssim 36$ pc) components, coincident with the two peaks imaged by Condon et al. Based on the Galactic Σ-D relation (which relates the surface brightness of Galactic SNRs to their diameter) and the ionizing flux from De Poy et al. (1986), Carral et al. concluded that there is far too much compact radio emission to be supported solely by SNRs, and that synchrotron emission from one or two active nuclei seems a more likely explanation.

We have made new radio observations of NGC 6240 with the Very Large Array (VLA) at 20 cm and 3.6 cm. These two new maps and four maps from previous VLA observations are used to perform a comprehensive study of the radio emission. In section 2, we describe the new observations and data reduction. Section 3 contains a discussion of the morphological features of the radio emission, based on the six radio maps. Spectral indices and intrinsic properties of the radio components are calculated. Possible mechanisms for producing the synchrotron emission from the different radio components are described in section 4.

## 2. Observations and Data Reduction

### 2.1. The Radio Maps

Two new observations of NGC 6240 have been made with the NRAO[2] VLA. On 1 September 1991, when the array was in 'A' configuration, the galaxy was observed at 8414.9 MHz (3.6 cm) for an effective integration time of 28 minutes. An additional deep exposure (135 minutes) was made on 1 November 1991 at 1489.9 GHz (20 cm) with the VLA in the 'B' configuration. Each observation had an effective bandwidth of 100 MHz. Partially processed data for four additional VLA observations were provided by previous workers (see Table 1).

[2]The National Radio Astronomy Observatory is operated by Associated Universities, Inc., under contract with the National Science Foundation.

The reduction of the radio data was performed with the NRAO AIPS data reduction package. Visibility data from the two new observations were flux-calibrated relative to 3C 286 and phase-calibrated using the nearby source 1648+015. Both observations had sufficient signal for self-calibration. After several iterations of CLEANing and self-calibration, the residual rms noise was 40 $\mu$Jy/beam and 100 $\mu$Jy/beam, yielding peak signal-to-noise (S/N) ratios of 390 and 1090 for the 3.6 and 20 cm images, respectively. Instrumental parameters, total fluxes and peak S/N ratios for the two new images (L and X1) are listed in Table 1, and contour plots are shown in Figure 2. Parameters of the four additional maps are listed in Table 1, and contour plots are shown in Figure 3.

Radio flux measurements of NGC 6240 from previous single-dish and pre-VLA interferometer observations are plotted in Figure 4. The total fluxes in each of the six VLA maps are listed in Table 1, and plotted in Figure 4 for comparison with the single-dish measurements. Total fluxes were computed by integrating the flux inside boxes enclosing the radio emission. Errors for the total flux were estimated as 5% for maps L, C1, C2, X1 and X2, and 10% for map U1. Note that the total fluxes in the low-resolution VLA maps (L, C2 and X2) are consistent with the single-dish measurements, indicating that little flux was missed by the VLA. On the other hand, the high-resolution maps (C1, X1 and U1) miss a significant fraction (40–60%) of the total radio flux from NGC 6240.

### 2.2. Method of Calculation of Spectral Indices

The group of low-resolution maps L, C2 and X2 are well-suited for a spectral index calculation, as are the group of high-resolution maps C1, X1, and U1 (see Table 1). Spectral indices at intermediate resolution were measured using maps C2 and X2. Special care was taken to ensure that the spectral index calculations would be reliable. In order to produce maps with matching coverage in the *u*-*v* plane, visibility data were truncated, when possible, at a common spacing at the lower end. Next, all maps, excepting the one with the largest beam size, were convolved with a Gaussian to give new maps with a common beam size. The convolved maps were then re-gridded so that each had the same pixel size, and the resulting images were aligned by peak pixel. Rectangular regions (Table 2) were then used to compute fluxes and spectral indices for each component (section 3.2).

## 3. Results

### 3.1. Radio Morphology

The new 20 cm map (Figure 2a) displays the strong, localized radio emission from the nuclear region as well as low surface brightness emission extending to the south and west of the nuclear source. Of the total flux, ∼68% emerges from within an $11'' \times 14''$ box centered on the nuclear radio source. Most of the remaining flux (∼26% of the total) is emitted by the arm extending to the west from the nuclear source. The 20 cm radio powers of the nuclear source and the western arm component ("N" and "W" in Figure 2a, respectively) are $\sim 10^{23}$ W $Hz^{-1}$ (Table 2). Here we use "N" to refer to the entire nuclear radio source, which includes the individual components N1, N2 and N3 (discussed later) plus the diffuse emission around them. Likewise, "W" refers to all of the the western arm clumps W1–W4 (see below) plus the emission between them.

The medium-resolution maps (Figures 3a and 3c) show that the western arm consists of several localized clumps of size $\sim 2{-}5''$ (1–2.5 kpc) plus fainter emission. These clumps are labeled W1, W2, W3 and W4 in Figures 3a and 3c. Fluxes and radio powers are listed in Table 2. The extension to the south from source N (Figure 2a) is resolved into a separate peak, and is labeled "S" in Figures 3a and 3c. Component S is located $\sim 7''$ south from source N and contributes $\sim 5\%$ to the total 6 cm flux.

The nuclear source N is resolved by the high-resolution maps (Figures 2b, 3b and 3d) into two compact sources (N1 and N2), plus a fainter source (N3). The radio peaks of the two nuclear sources N1 and N2 have a separation of $1.53''$ in the new 3.6 cm map (Figure 2b), along position angle (p.a.) $19.7°$. This value is in good agreement with those found in the radio images by Condon et al. (1982), Carral et al. (1990; Figure 3b), and Eales et al. (1990; Figure 3d). Source N3 is located $0.78''$ from N2 along p.a. $-67.3°$. Radio powers of components N1, N2 and N3 are listed in Table 2.

Each of the two nuclear sources N1 and N2 is composed of an unresolved component plus additional surrounding flux. By comparing the "peak flux" with the integrated "box flux" listed in Table 2, one notes that the unresolved component provides $\sim$60$-$70% of the flux in the composite sources N1 and N2. Hereafter we use the notation "N1$^*$" and "N2$^*$" when referring to the unresolved components of N1 and N2. A considerable fraction of the total flux in the high-resolution 3.6 cm map is from the diffuse component surrounding sources N1$^*$ and N2$^*$ (see Figure 2b). An additional diffuse component is implied by the difference of 7.4 mJy between the flux from source N in the medium-resolution 3.6 cm map and that in the high-resolution 3.6 cm map. Hereafter, we denote all of the diffuse emission in the nuclear region by "Nd": symbolically, N = N1$^*$ + N2$^*$ + N3 + Nd. The components N1$^*$, N2$^*$, N3 and Nd contribute 31%, 15%, 5% and 49% to the total 3.6 cm flux from source N.

### 3.2. Spectral Indices of the Radio Components

The spectral index $\alpha$ ($S_\nu \propto \nu^{-\alpha}$) of each of the individual and composite radio components was calculated using the method described in section 2.2. The results are listed in Table 3. The radio emission from the nuclear region of NGC 6240 (i.e. source N) has a spectral index of $\alpha \approx 0.6$. Source N3, which contributes only a small percentage of the total nuclear flux, has a steeper spectrum ($\alpha \approx 1.1$). The western arm W also has a steep spectrum ($\alpha \approx 1.0$). Three of the four components of the western arm have spectral indices comparable to that of the entire arm W. However, source W3, which emits ~25% of the total western arm flux, has a flatter spectrum ($\alpha \approx 0.7$). The spectral index for component S was found to be $\alpha \approx 0.6$.

In order to estimate the spectral index of the diffuse nuclear component Nd, we have calculated spectral indices for N1* and N2* from the peak flux in the rectangular regions (Table 2). Using the spectral indices of components N, N1*, N2* and N3 and the flux ratios for the components (at 3.6 cm; section 3.1), we estimate a value of $\alpha \approx 0.6$ for component Nd. We list values for the spectral index $\alpha$ and uncertainty $\Delta\alpha$ for components N1*, N2* and Nd in Table 3.

### 3.3. Intrinsic Properties of the Radio Components

The following properties of the radio components have been calculated: source radius, radio power, radio luminosity, total energy in particles and magnetic field, magnetic field strength, energy density in the magnetic field, total energy density in radiation, synchrotron and inverse-Compton (IC) lifetime of relativistic electrons and X-ray flux from IC scattering.

The radius $R$ of each component (Table 4, column 2) was measured from the radio maps. Spectral indices from Table 3 were used to calculate radio powers at 20 cm ($P_{20}$) and radio luminosities ($L$) over the frequency range 10 MHz – 100 GHz (Table 4, columns 3 and 4).

The magnetic field at minimum total energy can be calculated from (Pacholczyk 1970):

$$B_{(min)} = (4.5)^{2/7}(1+k)^{2/7}\phi^{-2/7}C_{\alpha}^{2/7}R^{-6/7}L^{2/7}.$$

Here $k$ is the ratio of the energy in heavy particles to that in electrons, $\phi$ is the volume filling factor for the relativistic particles and the magnetic field in the source, and $C_\alpha$ is a slowly-varying parameter (dependent on the spectral index and the limits of the radio window). We have taken $k = 1$ and $\phi = 1$. The minimum total energy in the particles and magnetic field may be calculated using the relation $E^{field}_{(min)} = \frac{3}{4}E^{part}_{(min)}$, which yields:

$$E^{tot}_{(min)} = E^{field}_{(min)} + E^{part}_{(min)} = \frac{7}{3}\frac{B^2_{(min)}}{8\pi}\frac{4}{3}\pi R^3\phi.$$

Values of the total energy $E^{tot}_{(min)}$, magnetic field $B_{(min)}$ and energy density in the magnetic field ($\epsilon_{MAG} = B^2_{(min)}/8\pi$) are listed in columns 5, 6 and 7 of Table 4, respectively. Changing the factor of $k$ to 100, as suggested by Burbidge (1959), will increase $E^{tot}_{(min)}$ by a factor of 9.4 and $B_{(min)}$ by a factor of 3.1. For the unresolved components N1* and N2*, the values obtained for $B_{(min)}$ and $\epsilon_{MAG}$ are lower limits, and that obtained for $E^{tot}_{(min)}$ is an upper limit.

In order to estimate the total energy density in radiation ($\epsilon_{RAD}$), we note that 70−80% of the 20$\mu$m flux arises within a 5.5″ beam centered on the nuclear region (De Poy et al. 1986). For our calculations, we have assumed that 75% of the total FIR luminosity originates from within source N. Although no additional information is available about the actual distribution of the FIR emission, we consider two exemplary cases.

For the first (I) case we assume that the FIR brightness distribution in the nuclear region is distributed as that of the radio emission. Using the high-resolution 3.6 cm radio map (Figure 2b) as a template, we estimate the structure of the FIR emission in the nuclear region as 3 sources at the locations of N1*, N2* and N3 (with luminosity 1.1, 0.5 and 0.2 × $10^{11}$ $L_\odot$, respectively) with the same angular diameters as the radio components, plus an additional component (luminosity $1.7 \times 10^{11}$ $L_{\odot}$) which is uniformly distributed within source N. We distribute the remaining 25% of the total FIR flux in five additional FIR sources at the locations of radio sources W1–W4 and S; the FIR flux was taken to be proportional to the radio flux from the individual components, measured from the medium-resolution 3.6 cm map (Figure 4c). Values of $\epsilon_{RAD}$ in a given radio component were then computed by summing the contributions from the assumed FIR sources. For the second (II) case, we also assume 75% of the FIR flux originates from the nuclear region. However, for this case we assume a uniform FIR brightness distribution within source N. The remaining 25% is assumed to be uniformly distributed out to a radius of 15″. For each case (I and II), we list values of $\epsilon_{RAD}$ in each radio component (Table 4, column 9).

The relativistic electrons in the radio sources suffer energy losses from synchrotron and IC radiation. The lifetime $\tau_{el}$ of the electrons is (cf. van der Laan & Perola 1969):

$$\tau_{el} \approx 1.0 \times 10^{3} \frac{B_{(min)}^{1/2}}{(\epsilon_{MAG} + \epsilon_{RAD})\nu^{1/2}} \quad yr,$$

where $B$ has units $G$, $\epsilon$ has units $erg\ cm^{-3}$, and $\nu$ is the frequency (in $Hz$) of the radio emission. In order to derive illustrative electron lifetimes, we take $\nu = 8.1$ GHz, which is the highest frequency at which fluxes exist for all radio components. The lifetime thus derived, $\tau_{el}$, is listed for each of the radio components (for cases I and II) in Table 4, column 10.

In either case, $\tau_{el}$ for the electrons in the nuclear sources is $\lesssim 3 \times 10^{4}$ yr, which is at least an order of magnitude smaller than that for the western arm sources. The relatively flat spectrum ($\alpha \approx 0.6$) of the nuclear emission suggests that the radio emitting electrons have not lost significant energy to synchrotron or IC scattering, in which case the relativistic electrons in the nuclear sources must be quite young ($\lesssim 3 \times 10^{4}$ yr). On the other hand, the western arm sources have steep ($\alpha \approx 1.0$) spectra at 8.1 GHz, suggesting ages in excess of $\sim 3 \times 10^{5}$ yr (Table 4).

The relativistic electrons producing the synchrotron radio emission also upscatter the FIR photons to X-ray and $\gamma$-ray energies. The energy loss from IC scattering is related to the energy loss in synchrotron radiation by (cf. Pacholczyk 1970):

$$\left(\frac{dE}{dt}\right)_{IC} = \frac{\epsilon_{RAD}}{\epsilon_{MAG}}\left(\frac{dE}{dt}\right)_{SYNC}.$$

Seed photons of wavelength $\lambda$ will produce IC radiation at the energy $E \approx (hc/\lambda)(\nu/\nu_g)$, where $\nu$ is the frequency of the synchrotron radiation and $\nu_g = eB/2\pi m_e c$ is the non-relativistic gyrofrequency. The spectrum of the IC radiation, which has the same form as that of the synchrotron radiation, can be written

$$F_E^{IC} \approx 1.4 \times 10^{-20}\left(\frac{\lambda}{60\ \mu\mathrm{m}}\right)\left(\frac{\mathrm{B}}{10^{-5}\ \mathrm{G}}\right)\left(\frac{S_\nu}{1\ \mathrm{mJy}}\right)\frac{\epsilon_{RAD}}{\epsilon_{MAG}}\ \ erg\ s^{-1}\ cm^{-2}\ keV^{-1},$$

where $S_\nu$ is the synchrotron flux at frequency $\nu = E\nu_g\lambda/hc$.

The relativistic electrons producing the radio emission at *cm* wavelengths scatter the FIR photons to energies of ~100 keV – 10 MeV. The IC flux at ~1 MeV is expected to be $\sim 10^{-17}$ erg s$^{-1}$ cm$^{-2}$ keV$^{-1}$, which is far too low to detect with existing instruments on board the CGRO. However, at lower X-ray energies (2–10 keV) the detectors are more sensitive. In order to compute the total IC flux ($F^{IC}$) from each radio component in this band, we have extrapolated the radio spectrum to *MHz* frequencies (relativistic electrons with $\gamma \sim 100$ are required) and have assumed that the FIR photons being upscattered have wavelength 60 $\mu$m. Values of $F^{IC}$ are shown (Table 4, column 11) for both cases of the assumed FIR radiation field.

The values of $\epsilon_{RAD}/\epsilon_{MAG}$ and $F^{IC}$ are highly dependent on the assumed radiation field. For case I, one can see from Table 4 that $\epsilon_{RAD}$ exceeds $\epsilon_{MAG}$ by factors of ~5 to >30. Consequently, for this case, IC losses dominate over synchrotron losses, especially within the nuclear components. On the other hand, for case II, $\epsilon_{RAD}$ is typically smaller than $\epsilon_{MAG}$ by a factor of a few, except in N3 and Nd, for which it is larger by factors of 3 and 30, respectively.

NGC 6240 has previously been observed in the 2–10 keV band with the HEAO–A1 instrument, but was undetected. The upper limit to the 2–10 keV flux is $4.2 \times 10^{-12}$ erg $s^{-1}$ $cm^{-2}$ (Rieke 1988). We predict an IC flux from all radio components of $\sim$2–6 $\times$ $10^{-14}$ erg $s^{-1}$ $cm^{-2}$, which is significantly lower than the HEAO-A1 limit. Significantly more IC radiation could be present if N1* and N2* contain ultra-compact ($R \lesssim$ several pc) radio and FIR cores (for case I: $F^{IC} \propto R^{-8/7}$).

The expected IC radiation in the soft (0.2–2.4 keV) X-ray band is of similar magnitude ($\sim 10^{-14}$ erg $s^{-1}$ $cm^{-2}$). The instruments on board the ASCA and ROSAT X-ray satellites should be able to detect emission at this flux level; however, it is possible that other components (e.g. AGN, starburst or hot, diffuse gas) may dominate the total X-ray emission.

## 4. Discussion

### 4.1. The Relation between Radio, Optical and Near Infrared Emission

Figure 5a shows the radio contours of the medium-resolution 6 cm map over a grayscale plot of a V-band image from Keel (1990). It is noteworthy that the individual western arm components are not correlated with the V-band emission, and are, in fact, located in regions of very weak V-band emission. We have also compared the 6 cm radio emission with a low-resolution K-band image (Figure 5b), also from Keel (1990). The K-band emission is much less affected by dust than the V-band emission, and thus better represents the distribution of stars in NGC 6240. The western arm is seen to correspond with an area of low K-band intensity. The lack of stellar light at the position of the western arm indicates that these radio sources are not spatially coincident with starburst activity.

In Figure 5c, the H$\alpha$ filaments described by Heckman et al. (1987;1990) are shown in a grayscale plot of an H$\alpha$ image from Keel (1990), again with the same radio contours. The strong filaments extending to the east and southeast are clearly noticeable, as are weaker filaments extending to the south and west. Some filaments seem to end at the radio components S, W1 and W4, suggesting that outflow from the nucleus may be related to these synchrotron sources.

The radio emission from source N does, however, emerge from an area of high V-band and K-band emission (Figures 5a and 5b), indicating a possible association with the stellar population. Figure 6 shows the radio contours of our new 3.6 cm map of the nuclear region overlayed on the high-resolution (sub-arcsecond) U-band HST image from Barbieri et al. (1993). The images were aligned using the relative positions (accuracy $0.2''$; H. Schulz, private communication) from Figure 2b of Barbieri et al. (1993). The resolved U-band emission from the two "nuclei" is not coincident with the radio emission, even when the uncertainty in the alignment is taken into account. Neither radio component N3 nor the diffuse radio emission surrounding N1$^*$ and N2$^*$ is correlated with the U-band emission. The emission visible in the U-band image is likely seen through lines of sight with particularly low obscuration. Therefore, the U-band image does not reflect the true distribution of starlight and the difference between the radio and optical morphology is probably dominated by obscuration effects. This interpretation is further supported by the fact that the separation of the double peaks increases monotonically as the wavelength decreases from the radio regime, through NIR, to optical (cf. Schulz et al. 1993), suggestive of an obscuring dust cloud concentrated between the peaks. Support for this hypothesis comes from the discovery that the $H_2$ line emission at 2 $\mu$m (Herbst et al. 1990; van der Werf et al. 1993) originates from between the two continuum peaks.

### 4.2. The Origin of the Radio Emission

The radio-FIR correlation implies that the mechanism for the production of the radio emission is closely connected with the process by which the dust is heated. We discuss interpretations of the nuclear radio emission in terms of SNRs and escaped electrons from the remnants, radio supernovae (RSNe) and an active nucleus. We then offer an interpretation of the western arm as radio emission from a galactic superwind. An alternative interpretation as a tidal feature is also considered, but not favored.

### *4.2.1. Supernova Remnants*

In the starburst scenario, the resulting SNe, SNRs and relativistic electrons which have leaked out of the SNRs are all potential sources of radio emission. In order to estimate how much radio emission should be produced, we must know the SN rate and we must make assumptions about the time evolution of the radio emission from the possible contributors. Previous workers have used the observed non-thermal radio emission to work backward and predict the SN rate (e.g. Ulvestad 1982; Condon & Yin 1990). One may estimate a SN rate $r$ from the non-thermal radio luminosity in a number of different ways, depending on which component is assumed to dominate the radio emission.

First, as a rough estimate, we assume that all of the radio emission comes from SNRs like Cas A, which has a radio lifetime of $\sim 10^4$ yr, a diameter of $\sim$3 pc, and $P_{20} = \sim 2 \times 10^{18}$ W Hz$^{-1}$. We can then estimate a filling factor for Cas A-like SNRs ($f_{CasA}$) and a SN rate ($r_{CasA}$) for each radio component (Table 5, columns 2 and 3).

A more popular method of calculating SN rates from radio luminosities has been to use the $\Sigma$-D relation (Clark & Caswell 1976), which relates the radio surface brightness and diameters of Galactic SNRs (e.g. Ulvestad 1982). We use the equation from Condon & Yin

(1990):

$$r_{\Sigma-D} = 7.1 \times 10^{-23} \left(\frac{408}{1489.9}\right)^{-\alpha} E_{50}^{1/17} n^{2/17} P_{20} \;\; yr^{-1},$$

where $E_{50}$ is the energy in the SN in units of $10^{50}$ erg, $n$ is the number density in the surrounding interstellar medium in cm$^{-3}$, $P_{20}$ is the 20 cm radio power in W Hz$^{-1}$, and $\alpha$ is the spectral index of the observed radio emission. Since $r$ changes very slowly with $E_{50}$ and $n$, we assume the typical values $E_{50}$ =1 and $n$ =1. We list a "$\Sigma$-D" SN rate ($r_{\Sigma-D}$) for each of the components in Table 5, column 4.

Condon & Yin (1990) have argued that the SN rate predicted from the $\Sigma$-D relation is a poor estimate of the true SN rate because the contribution to the radio emission from electrons after the adiabatic (Sedov) phase of a SNR is not included. In particular, the SN rate predicted from the $\Sigma$-D relation and the total radio power of the Milky Way is a factor of $\sim$20 larger than that estimated from counting SNRs (or from the pulsar birth rate). This implies that there is a contribution from the SNRs themselves after the adiabatic phase and/or from the relativistic electrons which escape from the SNRs. By scaling the non-thermal radio emission and SN rate from our Galaxy to the radio emission observed in NGC 6240, one can then calculate a SN rate (Condon & Yin 1990):

$$r_{C-Y} = 7.7 \times 10^{-24} (1.49)^{\alpha} P_{20} \;\; yr^{-1},$$

where $P_{20}$ is again the 20 cm radio power in W Hz$^{-1}$ and $\alpha$ is the spectral index of the observed radio emission. SN rates calculated using this method are listed in Table 5, column 5.

Supernova rates have also been estimated from stellar evolution models. Rieke et al. (1980) have successfully reproduced the FIR luminosity and other known observables in the starburst galaxies M82 and NGC 253 with starburst models. Rieke et al. (1985) performed computer simulations of the putative starburst in NGC 6240 and found that the SN rate scales with the FIR luminosity. Using these results, the SN rate in NGC 6240 would be 1.0–4.0 $yr^{-1}$. The radio powers of source N and W correspond to SN rates ($r_{C-Y}$) of 2.3 and 1.1 $yr^{-1}$, respectively. These values are consistent with those predicted by the independent simulations of Rieke et al., whereas the other two methods imply SN rates larger by an order of magnitude (Table 5).

However, special notice must be taken of the conditions in the two compact cores N1* and N2*. These sources have filling factors $f_{Cas\ A}$ of order unity, which means individual SNRs will overlap and form a hot bubble if they are older than ~300 yr (the age of Cas A). The calculation of the "Cas A" SN rate (column 3) assumes that the SNRs have the equivalent radio power of Cas A for $10^4$ yr, by which time the remnants will have merged. The Σ-D and Condon & Yin relations also assume that the radio emission comes from individual, separate SNRs. Thus, none of these methods for predicting the SN rate from the radio power are reliable for sources N1* and N2*.

If the SNRs are only ~300 yr old and barely overlap, extremely high SN rates are required. Each core emits the equivalent power of $\sim 3 \times 10^5$ Cas A's (Table 2). If the SNRs have the same radio power as Cas A, then the average SN rate over the past ~300 yr was $\sim 10^3$ $yr^{-1}$ per compact core. If, for example, the cores contain SNRs with radio power and radius equal to those values expected for Cas A at 100 years, a more reasonable SN rate of $r \sim 10$ $yr^{-1}$ per compact core is required. However, it is highly implausible that all of the SNe in the compact core are this young.

### *4.2.2. Radio Supernovae*

Massive stars in a starburst evolve into supernovae, which eventually become radio-loud in the remnant stage ($\gtrsim$ 100 yr). A number of supernovae have been known to emit strongly at radio wavelengths in the supernova stage ($\lesssim$ 10 yr), with radio powers 100–1000 times that of Cas A (see Sramek & Weiler 1990).

It has been proposed by Colina & Pérez-Olea (1992) that if the radio emission in ultra-luminous FIR galaxies is supported by a starburst, then the major contribution to the non-thermal radio emission is from RSNe. Although our knowledge of RSNe is limited, we investigate this possibility by assuming that the RSNe have the same radio power as the well-known RSN SN1979C ($P_{20} \approx 10^{20}$ W Hz$^{-1}$; Weiler et al. 1986). The nuclear cores N1* and N2* have radio luminosities of $\sim 5 \times 10^{22}$ W Hz$^{-1}$, so a group of ~500 RSNe could conceivably support the radio emission in the compact components. Columns 6 and 7 of Table 5 list the equivalent number of SN1979C's for the radio power of each component and the corresponding SN rate for a 10-yr radio lifetime. The SN rates predicted in this manner are two orders of magnitude higher than those predicted from the Condon & Yin method and from the Rieke et al. models, suggesting that RSNe do not dominate the radio emission from NGC 6240.

### *4.2.3. Active Nucleus*

As noted earlier, the radio flux from NGC 6240 is slightly larger than that predicted by the radio-FIR relation for spiral galaxies and starbursts. This may imply the presence of a component of radio emission additional to that powered by star formation. Indeed, De Poy et al. (1986) claim that the recombination line flux in NGC 6240 is too small for what one would expect from O and B stars in a super-starburst, and hence the primary source of heating for the dust must be an active nucleus. The compact radio morphology of the two nuclear cores, coupled with their high radio powers suggests that an active nucleus may be present.

Wilson & Willis (1980) have used double and triple radio source morphologies (reminiscent of the jet-plus-lobe radio morphologies observed in powerful radio galaxies) to argue for the presence of an active nucleus in Seyfert galaxies. The two nuclear sources

N1 and N2 could be interpreted in this way, but there is no evidence for radio emission from the putative nucleus in between them. A strong argument could be made if one or more ultra-compact (1 mas $\sim$ 0.5 pc) or flat-spectrum radio cores were found. Lonsdale et al. (1993) have conducted a search for ultra-compact radio cores in ultra-luminous FIR galaxies using VLBI techniques, and have found such cores in over half of the 31 galaxies in their sample. They suggest that these ultra-compact cores are associated with an active nucleus, which would imply that active nuclei are prevalent in luminous FIR galaxies such as NGC 6240. However, lacking milli-arcsecond VLBI observations of NGC 6240, one cannot judge whether N1* and N2* house ultra-compact radio cores associated with an active nucleus, or are powerful, compact starbursts.

### *4.2.4. Superwind*

As discussed in section 4.1, the clumps of radio emission in the western arm are located in a region of little starlight, and are coincident with the ends of H$\alpha$ filaments extending from the nucleus. Previous authors have interpreted radio emission from the halos of starburst galaxies as synchrotron emission from a "superwind" (a relativistic plasma driven out of the galaxy by supernovae from a powerful starburst; cf. Heckman, Lehnert & Armus 1993). Seaquist & Odegard (1991) have found that the spectrum of the radio emission in the nearby starburst galaxy M82 steepens from the nucleus to the halo. They attribute this spectral steepening to IC scattering by relativistic electrons on FIR photons from the starburst. In NGC 6240, we have found that the radio spectrum of the western arm is steeper than that of the nuclear sources. We therefore discuss an interpretation of the western arm as synchrotron emission from a superwind.

Observationally, the superwind manifests itself in the optical regime as H$\alpha$ filaments extending away from the nucleus, and in the radio regime as a steepening in the spectrum with radius from the galaxy center. The conditions necessary to produce a superwind are (i) there must be sufficient energy in the wind to eject the interstellar medium from the starburst region and (ii) the expansion time must be significantly shorter than the radiative cooling timescale (Heckman et al. 1993).

Assuming the wind to be starburst-driven, the first condition would be met if the volume filling factor of SNRs $f$ is near unity (Heckman et al. 1993). We have found filling factors in the two compact cores N1* and N2* to be of order unity (section 4.2.1). It is clear that densely-packed SNe in the two compact cores would indeed supply adequate power to drive a superwind away from the nuclear region. The first condition might also be satisfied if an AGN were present in the nucleus.

Heckman et al. (1990) have noted very broad optical line widths (~1000 km $s^{-1}$ FWHM) from the H$\alpha$ filaments extending outward from the nuclear region. This may be used to argue that nuclear outflow is occurring. A superwind that is blowing away from the nuclear region with an average velocity of 1000 km $s^{-1}$ would take $\sim 2.7 \times 10^6$ yr to reach the western arm (~2.8 kpc). Measurements of the central pressure in NGC 6240 imply a much longer timescale for radiative cooling ($\sim 4 \times 10^8$ yr; Heckman et al. 1990). Therefore the two conditions for producing a superwind are satisfied.

The relativistic electrons in the wind are assumed to have originated from the nuclear radio cores. We expect the radio spectrum of the wind to steepen from synchrotron and IC losses as it blows outward from the nuclear region. The steep-spectrum nuclear source N3 could possibly be a "clump" of ejected electrons which has not yet left the nuclear region. If N3 was ejected from N2*, it is travelling toward the western arm clump W1.

A superwind is expected to expand perpendicular to the disk in a spiral galaxy. An observational manifestation of this effect is the presence of H$\alpha$ filaments extending perpendicular to the disks in edge-on spirals (cf. Heckman et al. 1993). The major axis of the disk in NGC 6240 can be inferred from the K-band image (Figure 5b). The morphology of the western radio arm suggests the outflow in NGC 6240 is directed along p.a. $\approx -75°$, which is approximately perpendicular to the major axis. This is in agreement with the superwind outflow axis suggested by Heckman et al. (1990).

#### *4.2.5. The Western Arm as a Tidal Feature*

The western radio arm might alternatively be interpreted as a tidal feature. Simulations of interacting galaxies usually display such clumpy structure (cf. Barnes & Hernquist 1992), but it is unclear why the clumps would be powerful radio emitters. The radio power of the western arm ($P_{20} \sim 10^{23}$ W Hz$^{-1}$) is much higher than is typical of interacting spirals (Hummel 1981), and in the few systems which have comparable radio power, the observed radio emission may be powered by an active nucleus (e.g. NGC 4410a/b; Hummel, Kotanyi & van Gorkom 1986). A strong argument against the interpretation of the western radio arm as a tidal feature is the absence of associated starlight (section 4.1).

## 5. Conclusions

The diffuse radio emission in the nuclear region requires a supernova rate of 1.0 yr$^{-1}$ (if powered by supernova remnants), which is consistent with the results from the Rieke et al. models. If the radio emission in the two unresolved ($R \lesssim 36$ pc) radio cores N1$^*$ and N2$^*$ is powered by a starburst, then either the supernova remnants are extremely young ($\lesssim$100 yr), or the SNRs overlap and form hot bubbles, which could power a galactic superwind. The radio spectra of the individual nuclear components N1$^*$, N2$^*$ and Nd imply that these sources are very young ($\lesssim 3 \times 10^4$ yr).

No solid evidence for the existence of an active nucleus in NGC 6240 emerges from the VLA radio observations. VLBI observations of NGC 6240 could probe the structure of the two compact sources at milli-arcsecond resolution, which would provide further insight into the origin of the radio (and perhaps the far-infrared) emission in NGC 6240.

The lack of optical or near-infrared starlight coincident with the western arm radio sources implies that its radio emission is not generated by a local, off-nuclear starburst. The steep radio spectrum of the western arm suggests that the synchrotron-producing electrons have incurred significant synchrotron or inverse-Compton energy losses. Our interpretation of the western arm radio emission as synchrotron emission from a galactic superwind requires the presence of a powerful, compact starburst in the nuclear region or an active nucleus, either of which could power such a wind.

X-ray emission from inverse-Compton upscattered far-infrared photons should be present at detectable flux levels ($\sim$2–6 $\times$ $10^{-14}$ erg $s^{-1}$ $cm^{-2}$ in the 2–10 keV band). If the far-infrared radiation is concentrated within one or more ultra-compact ($\sim$ mas) cores associated with N1* and N2*, considerably more X-ray flux can be expected. X-ray observations with ROSAT and ASCA may be able to provide limits to the inverse-Compton flux, which could, in turn, provide information about the structure of the far-infrared sources in NGC 6240.

EJMC would like to thank Stephen White for helpful advice during the data reduction; Ski Antonucci, Patricia Carral, Gareth Wynn-Williams and Jun-Hui Zhao for sharing their VLA data; and Lee Armus, Cesare Barbieri, A. Baruffolo, Dave Davis, Tim Heckman, Tom Herbst, Mark Hereld, Bill Keel, Lee Mundy, Sachiko Okumura, Hartmut Schulz, Harley Thronson, Virginia Trimble, Zhong Wang, Martin Ward and Paul van der Werf for providing images and preprints and for offering helpful advice. This research was supported in part by NASA grants NAGW–2689 and NAGW–3268 to the Space Telescope Science Institute.

## Appendix

## "Dark Core" in NGC 6240?

No significant radio emission was found at the position of the possible "ultra-massive dark core" (Bland-Hawthorn, Wilson & Tully 1991) in either the new 3.6 cm or 20 cm map. The $3\sigma$ upper limit is 120 $\mu$Jy/beam at 3.6 cm. The large beam size of the 20 cm map smears the radio emission from the nuclear region of NGC 6240 out to the position of the dark core (Figure 2a). One can derive a $3\sigma$ upper limit to the 20 cm flux of 750 $\mu$Jy/beam from this map.

The upper limit to the 3.6 cm radio flux corresponds to a 20 cm radio luminosity $P_{20} < 4.8 \times 10^{20}$ W Hz$^{-1}$, assuming a spectral index of 0.7. The 20 cm radio luminosity function of Seyfert galaxies (Ulvestad & Wilson 1989, Figure 11) reveals that approximately 30 percent of such galaxies have 20 cm radio luminosities below that value. The absence of radio emission does not support the existence of a galactic nucleus at the supposed "dark core," but a radio-weak AGN is not ruled out.

Table 1: VLA Maps of NGC 6240

| Image Name | Figure No. | Freq (MHz) | VLA Config | Beam Size (″) | Beam P.A. (deg) | $u$-$v$ Range min (k$\lambda$) | $u$-$v$ Range max (k$\lambda$) | Total Flux (mJy) | S/N (peak/rms) | Ref |
|---|---|---|---|---|---|---|---|---|---|---|
| L | 2a | 1489.9 | B | 4.79×4.39 | -36.9 | 2.0 | 57 | 386 | 1090 | 1 |
| C1 | 3d | 4885.1 | A | 0.46×0.45 | -42.6 | 30 | 607 | 83.5 | 562 | 2 |
| C2 | 3a | 4585.1 | B | 2.17×1.59 | +0.4 | 2.0 | 350 | 131 | 477 | 3 |
| X1 | 2b | 8414.9 | A | 0.32×0.23 | -3.5 | 30 | 1010 | 53.4 | 390 | 1 |
| X2 | 3c | 8106.4 | C | 2.79×2.71 | +42.6 | 2.0 | 87 | 99.4 | 1850 | 4 |
| U1 | 3b | 14964.9 | A | 0.15×0.14 | -41.7 | 34 | 1820 | 33.1 | 171 | 5 |

References: (1) This paper; (2) Eales et al. 1990; (3) Antonucci 1985; (4) Zhao et al. 1993; (5) Carral et al. 1990

Table 2: Fluxes and Powers of Individual Radio Components

| Comp Name (1) | Image Used[1] (2) | Freq (MHz) (3) | Box Region (″) (4) | Box Flux[2] (mJy) (5) | Box Peak (mJy/beam) (6) | Power[3] ($10^{22}$ W Hz$^{-1}$) (7) | Power[3] ($10^3$ Cas A's) (8) |
|---|---|---|---|---|---|---|---|
| N | L | 1489.9 | 10.8 × 13.6 | 262±13 | 173 | 31 | 155 |
| W | L | 1489.9 | 12.0 × 14.8 | 100±15 | 35 | 12 | 60 |
| TOTAL | L | 1489.9 | | 386 | | 46 | 230 |
| N | C2 | 4585.1 | 5.2 × 6.0 | 94±5 | 58 | 11.3 | 126 |
| W1 | C2 | 4585.1 | 5.2 × 4.0 | 7.4±0.7 | 3.1 | 0.89 | 10 |
| W2[4] | C2 | 4585.1 | 3.2 × 5.2 | 9.3±0.7[4] | 4.0 | 1.11 | 12 |
| W3[4] | C2 | 4585.1 | 3.6 × 3.6 | 7.9±0.6[4] | 3.3 | 0.95 | 11 |
| W4[4] | C2 | 4585.1 | 6.0 × 6.4 | 7±1[4] | 2.7 | 0.8 | 9 |
| S | C2 | 4585.1 | 7.2 × 6.0 | 7±1 | 2.8 | 0.8 | 9 |
| TOTAL | C2 | 4585.1 | | 131 | | 15.7 | 174 |
| N1 | X1 | 8414.9 | 1.25 × 1.25 | 31±2 | 19 | 3.7 | 62 |
| N2 | X1 | 8414.9 | 1.0 × 1.0 | 14.4±0.7 | 9.1 | 1.7 | 28 |
| N3 | X1 | 8414.9 | 0.9 × 1.05 | 3.2±0.6 | 1.1 | 0.38 | 6.3 |
| TOTAL | X1 | 8414.9 | | 53.4 | | 6.4 | 107 |

[1] See Table 1

[2] Uncertainties for fluxes are calculated from an estimated 5% calibration error and the rms noise in the box regions.

[3] Radio powers were derived from the integrated flux in the box regions, using D = 100 Mpc ($H_o$ = 75 km s$^{-1}$ Mpc$^{-1}$). For comparison, the radio power of Cas A is 20, 9, and 6 × $10^{17}$ W Hz$^{-1}$ at 1489.9, 4585.1 and 8414.9 MHz, respectively.

[4] Component is not clearly separated from neighboring component; therefore, the actual uncertainty in the flux value is larger than that listed.

Table 3: Spectral Indices of Radio Components

| Comp Name | Resolution ($''$) | Data at Frequencies (GHz) | $\alpha$ (see note) |
|---|---|---|---|
| N | 4.6 | 1.5, 4.6, 8.1 | 0.65±0.07 |
| W | 4.6 | 1.5, 4.6, 8.1 | 1.01±0.07 |
| N | 2.75 | 4.6, 8.1 | 0.6±0.1 |
| W1 | 2.75 | 4.6, 8.1 | 1.0±0.1 |
| W2 | 2.75 | 4.6, 8.1 | 0.9±0.1 |
| W3 | 2.75 | 4.6, 8.1 | 0.7±0.1 |
| W4 | 2.75 | 4.6, 8.1 | 1.0±0.2 |
| S | 2.75 | 4.6, 8.1 | 0.6±0.1 |
| N1 | 0.45 | 4.9, 8.4, 15.0 | 0.70±0.15 |
| N2 | 0.45 | 4.9, 8.4, 15.0 | 0.80±0.15 |
| N3 | 0.45 | 4.9, 8.4 | 1.1±0.3 |
| N1* | 0.45 | 4.9, 8.4, 15.0 | 0.6±0.1 |
| N2* | 0.45 | 4.9, 8.4, 15.0 | 0.7±0.1 |
| Nd | ... | (see section 3.2) | 0.6±0.2 |

For the computations at medium-resolution ($2.75''$; and also for source N3), values for $\alpha$ and $\Delta\alpha$ were calculated directly from fluxes and errors in fluxes at the two frequencies listed. Otherwise (three frequencies available), $\alpha$ is the value at minimum-$\chi^2$ for a fit to a simple power-law, and $\Delta\alpha$ corresponds to $\Delta\chi^2 = 2.7$ (90% confidence).

Table 4. Properties of the Radio Components

| Name | Radius[1] | $P_{20}$[2] | Radio Lum[3] | $E^{tot}_{(min)}$ | $B_{(min)}$ | $\epsilon_{MAG}$ | Case[4] | $\epsilon_{RAD}$ | $\tau_{el}$ | $F^{IC,5}$ |
|---|---|---|---|---|---|---|---|---|---|---|
| | | $10^{22}$ | $10^{39}$ | $10^{54}$ | $10^{-5}$ | $10^{-10}$ | | $10^{-10}$ | $10^{5}$ | $10^{-15}$ |
| | $kpc$ | $W\ Hz^{-1}$ | $erg\ s^{-1}$ | $erg$ | $G$ | $erg\ cm^{-3}$ | | $erg\ cm^{-3}$ | $yr$ | $erg\ s^{-1}\ cm^{-2}$ |
| (1) | (2) | (3) | (4) | (5) | (6) | (7) | (8) | (9) | (10) | (11) |
| N1* | <0.04 | 6.3 | 12.6 | <0.1 | >44.0 | >76.9 | I | >2884 | <0.008 | >23.8 |
| | | | | | | | II | 21.7 | <0.2 | <0.2 |
| N2* | <0.04 | 3.9 | 6.1 | <0.1 | >41.0 | >66.9 | I | >1399 | <0.02 | >10.4 |
| | | | | | | | II | 21.7 | <0.3 | <0.2 |
| N3 | 0.1 | 2.6 | 1.8 | 0.3 | 15.3 | 9.4 | I | 58.8 | 0.2 | 4.3 |
| | | | | | | | II | 21.7 | 0.4 | 1.6 |
| Nd | 0.7 | 10.0 | 19.5 | 6.2 | 3.7 | 0.6 | I | 19.5 | 0.3 | 13.3 |
| | | | | | | | II | 21.7 | 0.3 | 14.8 |
| W1 | 0.9 | 2.7 | 2.5 | 5.2 | 2.5 | 0.2 | I | 1.5 | 3.2 | 3.4 |
| | | | | | | | II | 0.07 | 17.6 | 0.2 |
| W2 | 1.1 | 3.1 | 3.5 | 6.7 | 2.1 | 0.2 | I | 1.2 | 3.8 | 2.5 |
| | | | | | | | II | 0.07 | 20.4 | 0.2 |
| W3 | 0.9 | 2.1 | 3.4 | 3.6 | 2.1 | 0.2 | I | 1.1 | 3.9 | 0.7 |
| | | | | | | | II | 0.07 | 20.9 | 0.04 |

Table 4—Continued

| Name | Radius[1] | $P_{20}$[2] | Radio Lum[3] | $E^{tot}_{(min)}$ | $B_{(min)}$ | $\epsilon_{MAG}$ | Case[4] | $\epsilon_{RAD}$ | $\tau_{el}$ | $F^{IC,5}$ |
|---|---|---|---|---|---|---|---|---|---|---|
| | | $10^{22}$ | $10^{39}$ | $10^{54}$ | $10^{-5}$ | $10^{-10}$ | | $10^{-10}$ | $10^{5}$ | $10^{-15}$ |
| | *kpc* | $W\ Hz^{-1}$ | $erg\ s^{-1}$ | *erg* | *G* | $erg\ cm^{-3}$ | | $erg\ cm^{-3}$ | *yr* | $erg\ s^{-1}\ cm^{-2}$ |
| (1) | (2) | (3) | (4) | (5) | (6) | (7) | (8) | (9) | (10) | (11) |
| W4 | 0.9 | 2.6 | 2.4 | 5.0 | 2.4 | 0.2 | I | 1.3 | 3.5 | 3.0 |
| | | | | | | | II | 0.07 | 17.9 | 0.2 |
| S | 0.8 | 1.7 | 3.3 | 2.6 | 2.0 | 0.2 | I | 1.4 | 3.2 | 0.4 |
| | | | | | | | II | 0.07 | 21.1 | 0.02 |

[1] Component radii were measured using map U2 (Figure 3b; N1* and N2*), map X2 (Figure 2b; N3) and map C2 (Figure 3a; Nd, W1–W4 and S).

[2] Radio Powers at $\lambda = 20$ cm ($\nu = 1489.9$ MHz), derived from radio powers in Table 2 and the spectral indices in Table 3.

[3] Radio luminosities were derived by integrating the radio power from 10 MHz to 100 GHz. The radio emission was assumed to have the spectral index listed in Table 3 over the entire frequency range, except for the steep-spectrum sources (N3, W1, W2, and W4). For the steep-spectrum sources, the spectral index was taken to be $\alpha = 0.75$ from 10 MHz to $\nu_{cut}$, where $\nu_{cut}$ is the lowest frequency for which data exists for the component. From $\nu_{cut}$ to 100 GHz, the spectral index from Table 3 was used.

[4] Two cases (I and II) of FIR brightness distribution are explored (section 3.3).

[5] Flux from inverse Compton radiation over the energy range 2–10 keV.

Table 5: Supernova Rates and Filling Factors for the Radio Components

| Name | $f_{CasA}$ | $r_{CasA}$ | $r_{\Sigma-D}$ | $r_{C-Y}$ | $N_{SN1979C}$ | $r_{SN1979C}$ |
|---|---|---|---|---|---|---|
| | | $yr^{-1}$ | $yr^{-1}$ | $yr^{-1}$ | | $yr^{-1}$ |
| (1) | (2) | (3) | (4) | (5) | (6) | (7) |
| N1* | 2.6 | $(3.4)^{\dagger}$ | $(9.6)^{\dagger}$ | $(0.6)^{\dagger}$ | 630 | 63 |
| N2* | 1.6 | $(2.1)^{\dagger}$ | $(7.3)^{\dagger}$ | $(0.4)^{\dagger}$ | 393 | 39 |
| N3 | $4.1 \times 10^{-2}$ | 1.4 | 7.6 | 0.3 | 258 | 26 |
| Nd | $4.7 \times 10^{-4}$ | 5.3 | 15.4 | 1.0 | 998 | 100 |
| W1 | $6.6 \times 10^{-5}$ | 1.5 | 7.1 | 0.3 | 274 | 27 |
| W2 | $4.3 \times 10^{-5}$ | 1.6 | 7.0 | 0.3 | 307 | 31 |
| W3 | $5.0 \times 10^{-5}$ | 1.1 | 3.6 | 0.2 | 207 | 21 |
| W4 | $6.2 \times 10^{-5}$ | 1.4 | 6.7 | 0.3 | 259 | 26 |
| S | $5.6 \times 10^{-5}$ | 0.9 | 2.6 | 0.2 | 169 | 17 |

† The assumptions of the methods used for calculating these SN rates are violated in sources N1* and N2* (see text section 4.2.1).

---

This manuscript was prepared with the AAS LaTeX macros v3.0.

Fig. 1.— V-band image from Keel (1990), displayed on a logarithmic scale in order to bring out both strong and weak morphological features. North is up and east is to the left. The displayed $10''$ scale bar is 5 kpc for D = 100 Mpc. A dust lane is noticeable just to the west of the double peaks in the nuclear region. The distorted "tails" extending to the NE, SE and SW suggest that NGC 6240 may be the result of a recent merger between two galaxies.

Fig. 2.— New VLA maps of NGC 6240. **a)** 20 cm B-configuration map. The nuclear region is marked N and the western arm is marked W. The position of the suggested "dark core" is marked by a cross. Contour levels are -0.3 (-3$\sigma$), 0.3, 0.5, 1, 2, 6, 10, 14, 18, 22, 26, 30, 50, 70, 90, 110, 130 and 150 mJy/beam. The peak flux is 170.2 mJy/beam. **b)** 3.6 cm A-configuration map of the nuclear region. The size of this map is shown as a rectangle in Figure 2a. The double peaks are labeled N1 and N2, and an additional source to the NW of source N2 is labeled N3. Contour levels are -0.12 (-3$\sigma$), 0.12, 0.2, 0.4, 0.6, 0.8, 1, 2, 4, 6, 8, 10, 12, 14, and 16 mJy/beam. The peak flux is 17.6 mJy/beam. The size of the restoring beam (FWHM) is shown as an ellipse in the lower left corner of each map.

Fig. 3.— Additional VLA maps of NGC 6240. **a)** 6 cm B-configuration map from Antonucci (1985). The nuclear region is labeled N and the source to the south of the nuclear region is labeled S. The individual clumps in the western arm are labeled W1−W4. Contours are -0.25, 0.25, 0.5, 1.0, 1.5, 2.0, 2.5, 5, 15, 25, 35, 45 and 55 mJy/beam, and the peak flux is 57.9 mJy/beam. **b)** 2 cm A-configuration map of the nuclear region from Carral et al. (1990). The size of this map is shown as a rectangle in Figure 3a. The two compact cores that correspond with the two peaks in Figure 2b are labeled N1 and N2. Contours are -0.2, 0.2, 0.5, 1, 2, 4, 6, 8 and 10 mJy/beam. The peak flux is 10.7 mJy/beam. **c)** 3.6 cm C-configuration map from Zhao et al. (1993). The nuclear source N, southern source S, and western arm sources W1−W4 are labeled. Contours are -0.075, 0.075, 0.125, 0.25, 0.5, 0.75, 1, 2, 3, 4, 5, 10, 15, 20, 25, 30, 35, 40, 45 and 50 mJy/beam and the peak flux is 50.1 mJy/beam. **d)** 6 cm A-configuration map of the nuclear region from Eales et al. (1990). Again, the size of this map is shown as a rectangle in Figure 3c. Sources N1, N2 and N3 are labeled. Contours are -0.2, 0.2, 0.5, 1, 2, 4, 7, 10, 13, 16, 19, 22 and 25 mJy/beam. The peak flux is 28.9 mJy/beam. The sizes of the restoring beam (FWHM) are shown as ellipses in the lower left corner of each of the four maps.

Fig. 4.— The radio spectrum of NGC 6240. Previous single-dish (*open squares*) and pre-VLA interferometer flux measurements (*open circles*) were taken from Figure 2 of Fosbury & Wall (1979) at the following frequencies: 178, 318, 606, 2700, 5000, and 10600 MHz (single-dish); 408 and 1407 MHz (Cambridge One-Mile telescope); 962 MHz, (Jodrell Bank interferometer); and 2695 MHz (Cambridge 5-km telescope). The interferometric flux measurements at 408, 962 and 2695 MHz have missed a significant fraction of the total flux (cf. Fosbury & Wall 1979). The filled circles represent the total flux in the six VLA maps (L, C1, C2, X1, X2 and U1 in Table 1). The dashed line represents the best power-law fit to the pre-VLA measurements from 1407 MHz to 10600 MHz (which covers the frequency range of the six VLA maps).

Fig. 5.— Overlays of radio emission (contours) on optical and NIR images. **a)** 6 cm radio contours over a linearly-scaled grayscale plot of a V-band image from Keel (1990). The radio contours are identical to those in Figure 3a. The western arm of radio emission extends out from the nuclear region across the dust lane to regions of weak V-band emission. **b)** Same 6 cm radio contours over a logarithmically-scaled grayscale plot of a low-resolution K-band image from Keel (1990). The western arm is noted to extend over regions of weak K-band emission. **c)** Same 6 cm radio contours over a logarithmically-scaled grayscale plot of an H$\alpha$ image from Keel (1990). The H$\alpha$ filaments extending outward to the NW, SW and S, are directed toward the radio components W1, W3/W4 and S, respectively.

Fig. 6.— Overlay of 3.6 cm radio emission (contours) from the nuclear region on a logarithmically-scaled grayscale plot of an HST U-band image from Barbieri et al. (1993). The radio contours are the same as in Figure 2b. The separation between the double peaks is greater in the U-band than in the radio by $\approx 1''$, and there is little coincidence between the extended U-band emission and the extended radio emission.